\documentclass[12pt]{article}

\usepackage{amsmath}
\usepackage{amssymb}
\usepackage{amsfonts}
\usepackage{latexsym}
\usepackage{color}

\catcode `\@=11 \@addtoreset{equation}{section}

\catcode `\@=12

\newcommand{\be}{\begin{equation}}
\newcommand{\en}{\end{equation}}
\newcommand{\bea}{\begin{eqnarray}}
\newcommand{\ena}{\end{eqnarray}}
\newcommand{\beano}{\begin{eqnarray*}}
\newcommand{\enano}{\end{eqnarray*}}
\newcommand{\bee}{\begin{enumerate}}
\newcommand{\ene}{\end{enumerate}}

\newcommand{\mc}{\mathcal}

\newcommand{\D}{{\mc D}}

\newcommand{\Sc}{{\cal S}}
\newcommand{\E}{{\cal E}}
\newcommand{\F}{{\cal F}}
\newcommand{\G}{{\cal G}}

\newcommand{\Lc}{{\cal L}}

\newcommand{\C}{{\cal C}}
\newcommand{\1}{1 \!\! 1}

\newcommand{\Hil}{\mc H}

\newtheorem{thm}{Theorem}

\newtheorem{prop}[thm]{Proposition}
\newtheorem{defn}[thm]{Definition}

\catcode `\@=11 \@addtoreset{equation}{section}
\catcode `\@=12

\begin{document}

\thispagestyle{empty}

\vspace*{2cm}

\begin{center}
{\Large \bf Weak pseudo-bosons}   \vspace{2cm}\\

{\large F. Bagarello}\\
  Dipartimento di Ingegneria,
Universit\`a di Palermo,\\ I-90128  Palermo, Italy\\
and I.N.F.N., Sezione di Napoli\\
e-mail: fabio.bagarello@unipa.it\\
home page: www1.unipa.it/fabio.bagarello

\end{center}

\vspace*{2cm}

\begin{abstract}
\noindent We show how the notion of {\em pseudo-bosons}, originally introduced as operators acting on some Hilbert space, can be extended to a distributional settings. In doing so, we are able to construct a rather general framework to deal with generalized eigenvectors of the multiplication and of the derivation operators. Connections with the quantum damped harmonic oscillator are also briefly considered.
\end{abstract}

\vspace{2cm}


\vfill


\newpage

\section{Introduction}

In recent years pseudo-bosonic operators, \cite{baginbagbook}, have been used to rewrite some non self-adjoint Hamiltonian $H$ in terms of generalized (and, again, non self-adjoint) number operators. In this way, eigenvalues and eigenvectors of $H$, and of its adjoint $H^\dagger$, can be constructed using ladder operators acting on two different non zero vectors of the Hilbert space, the vacua of the theory.  Some applications of these operators can be found in references \cite{baginbagbook}-\cite{jon}. Pseudo-bosons are defined as operators acting on some dense domain in a given Hilbert space. Recently, in connection with the quantum damped harmonic oscillator, \cite{fff}, it has been shown that pseudo-bosons can be also used, at least formally, to diagonalize the Bateman Hamiltonian $H_B$ of the system, similarly to what other authors have done along the years. The point is that, being $H_B$ non bounded from below, the existence of a {\em real} vacuum (i.e., a vector in the Hilbert space) is not guaranteed at all. In fact, in \cite{fff} no such a vector is found, contrarily to what the authors in \cite{nakano} and \cite{deguchi} claim. On the other hand, in \cite{fff} it is shown that a vacuum of the annihilation pseudo-bosonic operator can only be found in a distributional sense. This motivates our present analysis: in this paper we will discuss what happens if two operators $a$ and $b$, originally defined on an Hilbert space $\Hil$ satisfies, in a suitable sense, the commutation relation $[a,b]=\1$. Here $\1$ is the identity operator on $\Hil$. Notice that this situation extends that of ordinary bosons, which is recovered if $b=a^\dagger$, and that of pseudo-bosons, for which it is assumed that a dense subspace of $\Hil$ exists, $\D$, which is left invariant by the action of $a$, $b$, and of their (Hilbert-)adjoints $a^\dagger$ and $b^\dagger$. In this paper the relevant aspect is that this set $\D\subset\Hil$ is replaced by a set of distributions, and that the biorthogonality of the eigenstates of the two adjoint number operators, see \cite{baginbagbook}, will be replaced by a {\em weak biorthogonality}, i.e. by a biorthogonality between distributions which, of course, should be defined properly.

The paper is organized as follows: in Section \ref{sectpbs} we review some standard results for {\em ordinary} $\D$-pseudo bosons.  In Section \ref{sectafa} we discuss a simple quantum mechanical system which, thought involving formal $\D$-pseudo bosons, does not satisfy any of the standard assumptions discussed in Section \ref{sectpbs}. 
 In Section \ref{sectwpbs} we introduce the notion of weak pseudo-bosons (wpbs in the following) and we study some of their properties. Section \ref{sectwpbs2} is devoted to  some preliminary considerations on the connections between wpbs and quantum damped harmonic oscillators. Our conclusions are given in Section \ref{sectconcl}.

\section{$\D$-pseudo bosons}\label{sectpbs}

In this section we briefly review some known facts on $\D$-pseudo bosons, to fix the notation and to better put later in evidence their differences with the wpbs.

Let $\Hil$ be a given Hilbert space with scalar product $\left<.,.\right>$ and related norm $\|.\|$. Let $a$ and $b$ be two operators
on $\Hil$, with domains $D(a)\subset \Hil$ and $D(b)\subset \Hil$ respectively, $a^\dagger$ and $b^\dagger$ their adjoint, and let $\D$ be a dense subspace of $\Hil$
such that $a^\sharp\D\subseteq\D$ and $b^\sharp\D\subseteq\D$, where with $x^\sharp$ we indicate $x$ or $x^\dagger$. Of course, $\D\subseteq D(a^\sharp)$
and $\D\subseteq D(b^\sharp)$.

\begin{defn}\label{def21}
The operators $(a,b)$ are $\D$-pseudo bosonic  if, for all $f\in\D$, we have
\be
a\,b\,f-b\,a\,f=f.
\label{A1}\en
\end{defn}

When $b=a^\dagger$, this is simply the canonical commutation relation (CCR) for ordinary bosons. However, when the CCR is replaced by (\ref{A1}), the situation changes, becoming mathematically more interesting. In particular, it is useful to assume the following:

\vspace{2mm}

{\bf Assumption $\D$-pb 1.--}  there exists a non-zero $\varphi_{ 0}\in\D$ such that $a\,\varphi_{ 0}=0$.

\vspace{1mm}

{\bf Assumption $\D$-pb 2.--}  there exists a non-zero $\Psi_{ 0}\in\D$ such that $b^\dagger\,\Psi_{ 0}=0$.

\vspace{2mm}

It is obvious that, since $\D$ is stable under the action of $b$ and $a^\dagger$, in particular,  $\varphi_0\in D^\infty(b):=\cap_{k\geq0}D(b^k)$ and  $\Psi_0\in D^\infty(a^\dagger)$, so
that the vectors \be \varphi_n:=\frac{1}{\sqrt{n!}}\,b^n\varphi_0,\qquad \Psi_n:=\frac{1}{\sqrt{n!}}\,{a^\dagger}^n\Psi_0, \label{A2}\en
$n\geq0$, can be defined and they all belong to $\D$. Then, they also belong to the domains of $a^\sharp$, $b^\sharp$ and $N^\sharp$, where $N=ba$. We see that, from a practical point of view, $\D$ is the natural space to work with and, in this sense, it is even more relevant than $\Hil$. Let's put $\F_\Psi=\{\Psi_{ n}, \,n\geq0\}$ and
$\F_\varphi=\{\varphi_{ n}, \,n\geq0\}$.
It is  simple to deduce the following lowering and raising relations:
\be
\left\{
    \begin{array}{ll}
b\,\varphi_n=\sqrt{n+1}\varphi_{n+1}, \qquad\qquad\quad\,\, n\geq 0,\\
a\,\varphi_0=0,\quad a\varphi_n=\sqrt{n}\,\varphi_{n-1}, \qquad\,\, n\geq 1,\\
a^\dagger\Psi_n=\sqrt{n+1}\Psi_{n+1}, \qquad\qquad\quad\, n\geq 0,\\
b^\dagger\Psi_0=0,\quad b^\dagger\Psi_n=\sqrt{n}\,\Psi_{n-1}, \qquad n\geq 1,\\
       \end{array}
        \right.
\label{A3}\en as well as the eigenvalue equations $N\varphi_n=n\varphi_n$ and  $N^\dagger\Psi_n=n\Psi_n$, $n\geq0$. In particular, as a consequence
of these last two equations,  if we choose the normalization of $\varphi_0$ and $\Psi_0$ in such a way $\left<\varphi_0,\Psi_0\right>=1$, we deduce that
\be \left<\varphi_n,\Psi_m\right>=\delta_{n,m}, \label{A4}\en
 for all $n, m\geq0$. Hence $\F_\Psi$ and $\F_\varphi$ are biorthogonal. It is easy to see that, if $b=a^\dagger$, then $\varphi_n=\Psi_n$, so that biorthogonality is replaced by a simpler orthonormality. Moreover, the relations in (\ref{A3}) collapse, and only one number operator exists, since in this case $N=N^\dagger$.

 The analogy with ordinary bosons suggests us to consider the following:

\vspace{2mm}

{\bf Assumption $\D$-pb 3.--}  $\F_\varphi$ is a basis for $\Hil$.

\vspace{1mm}

This is equivalent to requiring that $\F_\Psi$ is a basis for $\Hil$ as well, \cite{chri}. However, several  physical models show that $\F_\varphi$ is {\bf not} a basis for $\Hil$, but it is still complete in $\Hil$. For this reason we adopt the following weaker version of  Assumption $\D$-pb 3, \cite{baginbagbook}:

\vspace{2mm}

{\bf Assumption $\D$-pbw 3.--}  For some subspace $\G$ dense in $\Hil$, $\F_\varphi$ and $\F_\Psi$ are $\G$-quasi bases.

\vspace{2mm}
This means that, for all $f$ and $g$ in $\G$,
\be
\left<f,g\right>=\sum_{n\geq0}\left<f,\varphi_n\right>\left<\Psi_n,g\right>=\sum_{n\geq0}\left<f,\Psi_n\right>\left<\varphi_n,g\right>,
\label{A4b}
\en
which can be seen as a weak form of the resolution of the identity, restricted to $\G$. Of course, if $f\in\G$ is orthogonal to all the $\varphi_n$'s, or to all the $\Psi_n$'s, then (\ref{A4b}) implies that $f=0$. Hence $\F_\varphi$ and $\F_\Psi$ are complete in $\G$, \cite{bagbell}.

The families $\F_\varphi$ and $\F_\Psi$ can be used to define two densely defined operators $S_\varphi$ and $S_\Psi$ via their
action respectively on  $\F_\Psi$ and $\F_\varphi$: \be
S_\varphi\Psi_{ n}=\varphi_{ n},\qquad
S_\Psi\varphi_{ n}=\Psi_{bf n}, \label{213}\en for all $ n$, which also imply that
$\Psi_{ n}=(S_\Psi\,S_\varphi)\Psi_{ n}$ and
$\varphi_{ n}=(S_\varphi \,S_\Psi)\varphi_{ n}$, for all
$ n$. Of course, these equalities can be extended to the linear spans of the $\varphi_n$'s, $\Lc_\varphi$, and of the $\Psi_n$'s, $\Lc_\Psi$. This means that, for instance, $S_\Psi\,S_\varphi f=f$ and $S_\varphi\,S_\Psi g=g$ for all $f\in\Lc_\Psi$ and $g\in\Lc_\varphi$. With a little abuse of language we could say that $S_\varphi$ is the inverse of $S_\psi$. In fact, this is not always so, due to the fact that the two operators are, in general, unbounded and defined on different dense subsets of $\Hil$. However, there exist conditions, \cite{baginbagbook}, in which we can prove that $S_\Psi=S_\varphi^{-1}$. This is the case, for instance, when  $\F_\varphi$ and $\F_\Psi$ are Riesz bases. Quite often one writes these operators in a bra-ket form:
\be S_\varphi=\sum_{ n}\,
|\varphi_{ n}><\varphi_{ n}|,\qquad S_\Psi=\sum_{ n}
\,|\Psi_{ n}><\Psi_{ n}|, \label{212}\en
where $\left(|f\left>\right<f|\right)g=\left<f,g\right>f$, for all $f,g\in\Hil$.
These expressions may likely be
only formal, since the series are not necessarily convergent in
the uniform topology, as it happens when the operators $S_\varphi$ and $S_\Psi$ are unbounded. Again, this is not what happens if $\F_\varphi$ and $\F_\Psi$ are Riesz bases. In this case, we call our $\D$-pseudo bosons {\em regular}.

We end this short review by noticing that $S_\varphi$ and $S_\psi$ give rise to interesting intertwining relations between $N$ and $N^\dagger$:
 \be S_\Psi\,N\,g=N^\dagger S_\Psi\,g \quad \mbox{ and }\quad
N\,S_\varphi\, f=S_\varphi\,N^\dagger\,f, \label{219}\en $f\in\Lc_\psi$ and $g\in\Lc_\varphi$. This is
related to the fact that the eigenvalues of, say, $N$ and $N^\dagger$,
coincide and that their eigenvectors are related by the operators
$S_\varphi$ and $S_\psi$, in agreement with what is known on
intertwining operators, \cite{intop,bagio1}.

Many more results and examples on $\D$-quasi bosons can be found in \cite{baginbagbook}.

\section{Weak pseudo bosons: a first appearance}\label{sectafa}

Let us consider the following operators defined on $\Hil=\Lc^2(\mathbb{R})$: $\hat xf(x)=xf(x)$, $(\hat Dg)(x)=g'(x)$, the derivative of $g(x)$, for all $f(x)\in D(\hat x)=\{h(x)\in\Lc^2(\mathbb{R}: xh(x)\in \Lc^2(\mathbb{R} \}$ and $g(x)\in D(\hat D)=\{h(x)\in\Lc^2(\mathbb{R}: h'(x)\in \Lc^2(\mathbb{R} \}$. Of course, the set of test functions $\Sc(\mathbb{R})$ is a subset of both sets above: $\Sc(\mathbb{R})\subset D(\hat x)\cap D(\hat D)$. The adjoints of $\hat x$ and $\hat D$ in $\Hil$ are  $\hat x^\dagger=\hat x$, $\hat D^\dagger=-\hat D$. We have $[D,x]f(x)=f(x)$, for all $f(x)\in\Sc(\mathbb{R})$. This suggests that $\hat x$ and $\hat D$ could be thought as $\Sc(\mathbb{R})$-pseudo bosons, since they satisfy Definition \ref{def21} and since $\Sc(\mathbb{R})$ is stable under their action, and the action of their adjoints. However, if we look for the vacua of $a=\hat D$ and $b=\hat x$, we easily find that $\varphi_0(x)=1$ and $\psi_0(x)=\delta(x)$, with a suitable choice of the normalizations. It is clear, therefore, that neither $\varphi_0(x)$ nor $\psi_0(x)$ belong to $\Sc(\mathbb{R})$. And, more than this, they not even belong to $\Lc^2(\mathbb{R})$. Nevertheless, it is interesting to see what can be recovered of the framework proposed in Section \ref{sectpbs}, or if it can be extended, and how.

First of all, let us check if equation (\ref{A2}) still makes some sense. We have
\be
\varphi_n(x)=\frac{b^n}{\sqrt{n!}}\,\varphi_0(x)=\frac{x^n}{\sqrt{n!}}, \qquad \psi_n(x)=\frac{(a^\dagger)^n}{\sqrt{n!}}\,\psi_0(x)=\frac{(-1)^n}{\sqrt{n!}}\,\delta^{(n)}(x),
\label{31}\en
for all $n=0,1,2,3,\ldots$. Here $\delta^{(n)}(x)$ is the n-th weak derivative of the Dirac delta function. We can check that $\varphi_n(x), \psi_n(x)\in \Sc'(\mathbb{R})$, the set of the tempered distributions, \cite{gel}, that is the continuous linear functional on $\Sc(\mathbb{R})$. This suggests to consider $a^\dagger$ and $b$ as linear operators acting on $\Sc'(\mathbb{R})$. This is possible since the action of $\hat x$ and $\hat D$ can be extended outside $\Lc^2(\mathbb{R})$, to $\Sc'(\mathbb{R})$, which is  stable under the action of these operators. In other words: $a, b, a^\dagger$ and $b^\dagger$ all map $\Sc'(\mathbb{R})$ into itself. For this reason, we can further extend the pseudo-bosonic commutation relation, originally defined as $[D,x]f(x)=f(x)$, for all $f(x)\in\Sc(\mathbb{R})$, to the space of tempered distributions:
\be[a,b]\varphi(x)=\varphi(x),
\label{32}\en
for all $\varphi(x)\in\Sc'(\mathbb{R})$.

From (\ref{31}) it is clear that $b$ and $a^\dagger$ act as raising operators, respectively on the sets $\F_{\varphi}=\{\varphi_n(x)\}$ and $\F_{\psi}=\{\psi_n(x)\}$:
\be
b\varphi_k(x)=\sqrt{k+1}\varphi_{k+1}(x), \qquad \qquad a^\dagger\psi_k(x)=\sqrt{k+1}\psi_{k+1}(x),
\label{33}\en
$k=0,1,2,3,\ldots$. Equation (\ref{32}) implies that $b^\dagger$ and $a$ act as lowering operators on these sets:
\be
a\varphi_k(x)=\sqrt{k}\varphi_{k-1}(x), \qquad \qquad b^\dagger\psi_k(x)=\sqrt{k}\psi_{k-1}(x),
\label{34}\en
$k=0,1,2,3,\ldots$, with the understanding that $a\varphi_0(x)=b^\dagger\psi_0(x)=0$. It is now clear that, calling $N=ba=\hat x \hat D$, $N\varphi_k(x)=k\varphi_k(x)$, for all $k=0,1,2,3,\ldots$. This is because $N\varphi_k(x)=b(a\varphi_k(x))=\sqrt{k}\,b\varphi_{k-1}(x)=k\varphi_k(x)$. But the same result can also be found directly:
$$
N\varphi_k(x)=\hat x\, \hat D\,\frac{x^k}{\sqrt{k!}}=\hat x\, \frac{kx^{k-1}}{\sqrt{k!}}= \frac{kx^{k}}{\sqrt{k!}}= k\varphi_k(x).
$$
The distributions $\psi_k(x)$ are also (generalized) eigenstates of a number-like operator. In fact, calling $N^\dagger=a^\dagger b^\dagger$, and using formulas (\ref{33}) and (\ref{34}), one proves that $N^\dagger \psi_k(x)=k\psi_k(x)$. Again, this can be checked explicitly by computing
$$
N^\dagger \psi_k(x)=-\hat D \hat x\frac{(-1)^n}{\sqrt{n!}}\,\delta^{(n)}(x)=(-1)^{n+1} (x\delta^{(n)}(x))'=n\psi_n(x),
$$
since the weak derivative of $x\delta^{(n)}(x)$ can be easily computed and we have $(x\delta^{(n)}(x))'=-n\delta^{(n)}(x)$ for all $n=0,1,2,3,\ldots$. Summarizing, we have deduced that
\be
N\varphi_k(x)=k\varphi_{k}(x), \qquad \qquad N^\dagger\psi_k(x)=k\psi_{k}(x),
\label{35}\en
for all $k=0,1,2,3,\ldots$. This formula, together with (\ref{33}) and (\ref{34}), reflect the analogous results listed in Section \ref{sectpbs} suggesting, therefore, that a similar framework can be extended from the Hilbert space $\Lc^2(\mathbb{R})$ to the set of tempered distributions. The next step consists in checking, if possible, the biorthogonality of the sets $\F_{\varphi}$ and $\F_{\psi}$, and their basis properties, if any. In other words, we are interested in understanding whether equations (\ref{A4}) and (\ref{A4b}), or some similar expressions, can be deduced for our families of tempered distributions.

First of all we notice that the biorthogonality of $\F_{\varphi}$ and $\F_{\psi}$ would be guaranteed if each $\varphi_n(x)$ and $\psi_n(x)$ were ordinary functions, at least if $N^\dagger$ is the {\em standard} adjoint of $N$. None of these requirements is satisfied here: we have already seen that $\varphi_n(x)$ and $\psi_n(x)$ do not belong to $\Lc^2(\mathbb{R})$, and $N^\dagger=a^\dagger b^\dagger$ is only the formal adjoint of $N$, due to the nature of our operators. Nevertheless, as we will show next, $\F_{\varphi}$ and $\F_{\psi}$ are, indeed, biorthogonal. This property, in fact, can be deduce from what is discussed in some papers and books dealing with distributions, see \cite{morton,kanwal,estrada} for instance. Here we give our own proof, which is useful to fix, in our context and for future convenience, the general meaning of biorthogonality for distributions.

First we observe that the scalar product between two {\em good} functions, for instance $f(x),g(x)\in\Sc(\mathbb{R})$, can be written in terms of a convolution between $\overline{f(x)}$ and the function $\tilde{g}(x)=g(-x)$: $\left<f,g\right>=(\overline{f}* \tilde{g})(0)$. Following \cite{vlad}, in \cite{bag2017} this approach was used in a quantum mechanical settings, to extend the ordinary scalar product of $\Lc^2(\mathbb{R})$ to two Dirac delta functions. In the same way we define the scalar product between two elements $F(x), G(x)\in\Sc'(\mathbb{R})$ as the following convolution:
\be
\left<F,G\right>=(\overline{F}* \tilde{G})(0),
\label{36}\en
whenever this convolution exists. This existence issue is discussed in \cite{vlad}. As we will see, this will not be a problem for us. In order to compute $\left<F,G\right>$, it is necessary to compute $(\overline{F}* \tilde{G})(f)$, $f(x)\in\Sc(\mathbb{R})$, and this can be computed by using the equality\footnote{We stress once more that $(\overline{F}* \tilde{G})(f)$ is not always defined, but there exist useful situations when it is. This is the case when $\left<F,G*f\right>$ exists.} $(\overline{F}* \tilde{G})(f)=\left<F,G*f\right>$.

In our situation we have $F(x)=x^n$ and $G(x)=\delta^{(m)}(x)$, $n,m=0,1,2,3,\ldots$. Hence $(G*f)(x)=\int_{\mathbb{R}} \delta^{(m)}(y)f(x-y)dy=f^{(m)}(x)$, where $f^{(m)}(x)$ is the ordinary m-th derivative of the test function $f(x)$. Then we have
$$
(\overline{F}* \tilde{G})(f)=\left<F,G*f\right>=\int_{\mathbb{R}}\overline{F(x)}f^{(m)}(x)\,dx=\int_{\mathbb{R}} x^n\,\frac{d^m f(x)}{dx^m}\,dx=(-1)^m\int_{\mathbb{R}}\frac{d^m x^n}{dx^m}\,f(x)\,dx.
$$
Since 
$$
\frac{d^m x^n}{dx^m}=\left\{
\begin{array}{ll}
 0 \hspace{2.6cm} \mbox{if } m>n\\
 n! \hspace{2.5cm} \mbox{if } m=n\\
\frac{n!}{(n-m)!}\,x^{n-m}\hspace{0.8cm} \mbox{if } m<n,\\
\end{array}
\right.
$$
we conclude that 
$$
(\overline{F}* \tilde{G})(f)=\left\{
\begin{array}{ll}
0 \hspace{5.6cm} \mbox{if } m>n\\
(-1)^nn! \hspace{4.5cm} \mbox{if } m=n\\
(-1)^m\frac{n!}{(n-m)!}\,\int_{\mathbb{R}}x^{n-m}f(x)\,dx\hspace{0.8cm} \mbox{if } m<n,\\
\end{array}
\right.
$$
which implies that $(\overline{F}* \tilde{G})(0)=(-1)^n n! \delta_{n,m}$. Therefore,
\be\left<\varphi_n,\psi_m\right>=\delta_{n,m},
\label{37}\en
as claimed before. Notice that our original choice of normalization for $\varphi_0(x)$ and $\psi_0(x)$ guarantees the biorthonormality (and not only the biorthogonality) of the families $\F_\varphi$ and $\F_{\psi}$. 

\vspace{2mm}

It is clear that it makes no sense to check if $\F_\varphi$ or $\F_{\psi}$, or both, are bases or $\D$-quasi bases. This is because none of the $\varphi_n(x)$ and $\psi_n(x)$ even belongs to $\Lc^2(\mathbb{R})$. However, the pair $(\F_{\varphi},\F_{\psi})$ can still be used to expand a certain class of functions, those which admit expansion in Taylor series. In fact we have 
$$
\sum_{n=0}^\infty \left<\psi_n,f\right>\varphi_n(x)=\sum_{n=0}^\infty \frac{(-1)^n}{n!} \left<\delta^{(n)},f\right>x^n=\sum_{n=0}^\infty \frac{1}{n!} f^{(n)}(0)\,x^n=f(x),
$$
for all $f(x)$ admitting this kind of expansion. However, if we invert the role of $\F_{\psi}$ and $\F_{\varphi}$, the result is more complicated:
$$
\sum_{n=0}^\infty \left<\varphi_n,f\right>\psi_n(x)=\sum_{n=0}^\infty\frac{(-1)^n}{n!}\left<x^n,f\right>\delta^{(n)}(x).
$$
This is, in principle, an infinite series of derivatives of delta of the kind appearing in some literature on distributions, see \cite{kanwal,estrada} for instance. Series of this kind sometimes are called {\em dual Taylor series}. It is known that the series does not define in general an element of $D'(\mathbb{R})$, a distribution, (hence it cannot define a tempered distribution) except when the number of non zero moments of $f(x)$, $\left<x^n,f\right>$, is finite. In this case the series above returns a finite sum, and the result of this sum is indeed a tempered distribution.

This preliminary analysis shows that the pair $(\F_{\varphi},\F_{\psi})$ obeys what we can call {\em a weak  basis property}, but only for very special functions or distributions. What we will do next is to check if, and for which objects, a formula like the one in (\ref{A4b}) can be written. In this perspective, let us introduce the following set of functions:
\be
\D=\Lc^1(\mathbb{R})\cap \Lc^\infty(\mathbb{R})\cap A(\mathbb{R}),
\label{38}
\en
where $A(\mathbb{R})$ is the set of  entire real analytic functions, which admit expansion in Taylor series, convergent everywhere in $\mathbb{C}$. Of course $\D$ contains many functions of $\Sc(\mathbb{R})$, but not all.

Let now $f(x),g(x)\in\D$, and let us consider the following sequence of functions: $R_N(x)=\overline{f(x)}\,\sum_{n=0}^{N}\frac{g^{(n)}(0)}{n!}\,x^n$. It is clear, first of all, that $R_N(x)$ converges to $\overline{f(x)}\,g(x)$ almost everywhere (a.e.) in $\mathbb{R}$. Of course, it also converges with respect to stronger topologies, but this is not relevant for us. The second useful property is that $R_N(x)$ can be estimated as follows:
\be
|R_N(x)|\leq R(x)\equiv|f(x)|(M+\|g\|_\infty),
\label{addi}\en
for some fixed $M>0$ and for all $N$ large enough. It is clear that $R(x)\in\Lc^1(\mathbb{R})$. To prove the estimate in (\ref{addi}) it is enough to observe that, a.e. in $x$,
$$
|R_N(x)|\leq |f(x)|\left(\left|\sum_{n=0}^{N}\frac{g^{(n)}(0)}{n!}\,x^n-g(x)\right|+|g(x)|\right)\leq  |f(x)|(M+\|g\|_\infty),
$$
where $M$ surely exists (independently of $x$) due to the uniform convergence of $\sum_{n=0}^{N}\frac{g^{(n)}(0)}{n!}\,x^n$ to $g(x)$. Then we can apply the Lebesgue dominated convergence theorem, and therefore
$$
\lim_{N,\infty}\int_{\mathbb{R}} R_N(x)dx=\int_{\mathbb{R}} \overline{f(x)}\,g(x)dx=\left<f,g\right>.
$$
Incidentally we observe that, since $f,g\in\D$, $|\left<f,g\right>|\leq\|f\|_1\|g\|_\infty$, which ensures that $\left<f,g\right>$ is well defined. Now,
$$
\left<f,g\right>=\lim_{N,\infty}\int_{\mathbb{R}} R_N(x)dx=\sum_{n=0}^\infty\frac{1}{n!}\,g^{(n)}(0)\left<f,x^n\right>= \sum_{n=0}^\infty\frac{(-1)^n}{n!}\,\left<f,x^n\right>\left<\delta^{(n)},g\right>=
$$
$$
=\sum_{n=0}^\infty\,\left<f,\varphi_n\right>\left<\psi_n,g\right>.
$$
In a similar way we can also check that, for the same $f(x)$ and $g(x)$,
$$
\left<f,g\right>=\sum_{n=0}^\infty\,\left<f,\psi_n\right>\left<\varphi_n,g\right>.
$$
Hence we conclude that $(\F_{\varphi},\F_{\psi})$ are $\D$-quasi bases. It should be stressed that it is not clear if $\D$ is dense or not in $\Hil$, but this is not very relevant in the present context, where the role of the Hilbert space is only marginal. Moreover, there are also distributions which satisfy (half of) formula (\ref{A4b}). For instance, if $f(x)=\sum_{k=0}^{M}a_k\psi_{k}(x)$ for some complex $a_k$ and fixed $M$ the equality $\left<f,g\right>=\sum_{n=0}^\infty\,\left<f,\varphi_n\right>\left<\psi_n,g\right>$ is automatically satisfied, while it is not even clear that $\sum_{n=0}^\infty\,\left<f,\psi_n\right>\left<\varphi_n,g\right>$ is convergent. Similarly, if we take $g(x)=\sum_{k=0}^{L}b_k\varphi_{k}(x)$ for some complex $b_k$ and fixed $L$, $\left<f,g\right>=\sum_{n=0}^\infty\,\left<f,\varphi_n\right>\left<\psi_n,g\right>$ is true, while $\sum_{n=0}^\infty\,\left<f,\psi_n\right>\left<\varphi_n,g\right>$ could  be not even convergent.

In analogy with what we have done in Section \ref{sectpbs}, we can use $\F_{\varphi}$ and $\F_{\psi}$ to introduce two operators, $S_\varphi$ and $S_\psi$, which we formally write, for the moment, as in (\ref{212}):
\be S_\varphi=\sum_{ n}\,
|\varphi_{ n}><\varphi_{ n}|,\qquad S_\psi=\sum_{ n}
\,|\psi_{ n}><\psi_{ n}|. \label{39}\en
We have seen that these operators have interesting properties, and it makes sense to understand if they can be extended, and in which sense, to the present distributional context. 

First of all, we introduce the following subsets of $\Sc'(\mathbb{R})$:
$$
D(S_\varphi)=\{F(x)\in\Sc'(\mathbb{R}): \, (S_\varphi F)(x)\in \Sc'(\mathbb{R})\}
$$
and 
$$
D(S_\psi)=\{F(x)\in\Sc'(\mathbb{R}): \, (S_\psi F)(x)\in \Sc'(\mathbb{R})\}.
$$
We call these sets the {\em generalized domains} of $S_\varphi$ and $S_\psi$, respectively. It is easy to see that $\Lc_\varphi\subseteq D(S_\psi)$ and  $\Lc_\psi\subseteq D(S_\varphi)$ and that $S_\varphi: \Lc_\psi\rightarrow\Lc_\varphi$, while $S_\psi: \Lc_\varphi\rightarrow\Lc_\psi$. In particular we have
\be S_\varphi\left(\sum_{k=0}^Nc_k\psi_k\right)=\sum_{k=0}^Nc_k\varphi_k, \qquad S_\psi\left(\sum_{k=0}^Nc_k\varphi_k\right)=\sum_{k=0}^Nc_k\psi_k,
\label{310}\en
as well as
\be
S_\varphi S_\psi F=F, \qquad S_\psi S_\varphi G=G,
\label{311}\en
and
\be
NS_\varphi G=S_\varphi N^\dagger G, \qquad N^\dagger S_\psi F=S_\psi N F,
\label{312}\en
for $F(x)\in \Lc_\varphi$, $G(x)\in \Lc_\psi$. Furthermore, it is possible to see that $\Lc_\psi\neq D(S_\varphi)$. In fact, for $F$ to belong to $D(S_\varphi)$, it is sufficient that the series $\sum_{n=0}^\infty\left<\varphi_n,F\right>\varphi_n(x)=\sum_{n=0}^\infty\alpha_nx^n$, $\alpha_n=\frac{1}{n!}\left<x^n,F\right>$, converges. For instance, if $F(x)$ is equal to 1 for $x\in[0,1]$ and zero otherwise, the series converges for all $x\in\mathbb{R}$, even if $F(x)\notin\Lc_\psi$. 

We end this section by noticing that those in (\ref{312}) are not the only intertwining relations we can deduce for our operators. In fact, simple computations show that the following relations also hold:
$$
S_\psi a F=b^\dagger S_\psi F, \qquad S_\psi b F = a^\dagger S_\psi F,
$$
for all $F\in \Lc_\varphi$, and
$$
S_\varphi a^\dagger  G=b S_\varphi G, \qquad S_\varphi b^\dagger G = a S_\varphi G,
$$
for all $G\in \Lc_\psi$. These formulas can be rewritten as follows:
\be
a F=S_\varphi b^\dagger S_\psi F, \quad b F=S_\varphi a^\dagger S_\psi F, \qquad a^\dagger G=S_\psi b S_\varphi G, \quad  b^\dagger G=S_\psi a S_\varphi G, 
\label{313}\en
for all $F\in \Lc_\varphi$ and  $G\in \Lc_\psi$. In the language of \cite{bagJMP2013}, for instance, the operators $a$ and $b^\dagger$ could be called $S_\psi$-conjugate.

\section{Weak pseudo bosons: general settings}\label{sectwpbs}

Having in mind the results of the previous section, we introduce here a  definition which extends that of Section \ref{sectpbs}. First we consider two operators $a$ and $b$ which, together with their adjoints $a^\dagger$ and $b^\dagger$, map a certain dense subset of $\Hil$, $\D$, into itself. Further we assume that $a$ and $b$ can be extended to larger set, $\E\supset\Hil$, which is again stable under their action, and under the action of their adjoints.

\begin{defn}
	The operators $a$ and $b$ are {\em weak} $\E$-pseudo bosonic if 
	\be
	[a,b]\,F=F,
	\label{41}\en
	for all $F\in \E$. When the role of $\E$ is clear we will simply call $a$ and $b$  {\em weak} pseudo bosonic operators.
\end{defn}
As in Section \ref{sectpbs}, the commutator in (\ref{41}) is just the starting point to construct an interesting mathematical framework. This is exactly what we will do now, trying to rewrite the results deduced  in the previous section in an abstract form. The next two  assumptions reflect Assumptions $\D$-pb 1 and $\D$-pb 2:

\vspace{2mm}

{\bf Assumption $\E$-wpb 1.--}  there exists a non-zero $\varphi_{ 0}\in\E$ such that $a\,\varphi_{ 0}=0$.

\vspace{1mm}

{\bf Assumption $\E$-wpb 2.--}  there exists a non-zero $\Psi_{ 0}\in\E$ such that $b^\dagger\,\Psi_{ 0}=0$.

\vspace{2mm}

As before, the invariance of $\E$ under the action of the operators $a$, $b$, $a^\dagger$ and $b^\dagger$ implies that  $\varphi_0\in D^\infty(b):=\cap_{k\geq0}D(b^k)$ and  $\Psi_0\in D^\infty(a^\dagger)$, in the sense of generalized domains, so
that the vectors \be \varphi_n:=\frac{1}{\sqrt{n!}}\,b^n\varphi_0,\qquad \Psi_n:=\frac{1}{\sqrt{n!}}\,{a^\dagger}^n\Psi_0, \label{42}\en
$n\geq0$, can be defined and they all belong to $\E$. This is in fact what equation (\ref{31}) shows, identifying $\E$ with $\Sc'(\mathbb{R})$. Defining now the sets  $\F_\psi=\{\psi_{ n}, \,n\geq0\}$ and
$\F_\varphi=\{\varphi_{ n}, \,n\geq0\}$, from (\ref{41}) and from the definition in (\ref{42}) we easily deduce the same raising and lowering relations as in (\ref{A3}), together with the  eigenvalue equations
$N\varphi_n=n\varphi_n$ and  $N^\dagger\Psi_n=n\Psi_n$, $n\geq0$, where, once more, $N^\dagger$ is identified with $a^\dagger b^\dagger$. It is natural to assume now that, with a suitable choice of  the normalization of $\varphi_0$ and $\Psi_0$,  $\left<\varphi_0,\Psi_0\right>=1$, then
\be \left<\varphi_n,\Psi_m\right>=\delta_{n,m}, \label{43}\en
for all $n, m\geq0$. This means that $\F_\Psi$ and $\F_\varphi$ are requested to be biorthonormal, with respect to a bilinear form $\left<.,.\right>$ which extends the ordinary scalar product to $\E$. This is exactly what we deduced in Section \ref{sectafa},  where $\left<.,.\right>$ was extended to $\Sc'(\mathbb{R})$ using convolution of distributions.

Of course, it makes not much sense to require any strong version of the basis property for $\F_\Psi$ or $\F_\varphi$, in general. This is  evident from our results in Section \ref{sectafa}. What seems natural to require is that a set $\C\subseteq\E$ exists, sufficiently large\footnote{Of course, from a mathematical side, this is not really a {\em good} requirement, since we are not defining the set $\C$, but we are only requiring that it consists of a sufficiently large number of functions.}, such that
\be
\left<F,G\right>=\sum_{n=0}^\infty\,\left<F,\psi_n\right>\left<\varphi_n,G\right>=\sum_{n=0}^\infty\,\left<F,\varphi_n\right>\left<\psi_n,G\right>,
\label{44}\en
for all $F,G\in\C$. In particular, in the previous section,  $\C$ was identified with the set in (\ref{38}), which contains all the test functions which admit power expansion. A pragmatic view to $\C$ is that it should contains all those (generalized) functions which are somehow interesting for us. Of course, this will depend on the particular (physical or mathematical) system we are investigating.

As in Section \ref{sectafa}, we  use $\F_{\varphi}$ and $\F_{\psi}$ to introduce two operators, $S_\varphi$ and $S_\psi$, as follows:
let
$$
D(S_\varphi)=\{F\in\E: \, S_\varphi F\in \E\}, \qquad
D(S_\psi)=\{F(x)\in\E: \, S_\psi F\in \E\}.
$$
These are, see Section \ref{sectafa}, the { generalized domains} of $S_\varphi$ and $S_\psi$, respectively. All the properties found in the previous section are recovered: $\Lc_\varphi\subseteq D(S_\psi)$,  $\Lc_\psi\subseteq D(S_\varphi)$, $S_\varphi: \Lc_\psi\rightarrow\Lc_\varphi$, and $S_\psi: \Lc_\varphi\rightarrow\Lc_\psi$. In particular we have
\be S_\varphi\left(\sum_{k=0}^Nc_k\psi_k\right)=\sum_{k=0}^Nc_k\varphi_k, \qquad S_\psi\left(\sum_{k=0}^Nc_k\varphi_k\right)=\sum_{k=0}^Nc_k\psi_k,
\label{45}\en
as well as
\be
S_\varphi S_\psi F=F, \qquad S_\psi S_\varphi G=G,
\label{46}\en
and
\be
NS_\varphi G=S_\varphi N^\dagger G, \qquad N^\dagger S_\psi F=S_\psi N F.
\label{47}\en
Moreover
\be
a\, F=S_\varphi b^\dagger S_\psi F, \quad b\, F=S_\varphi a^\dagger S_\psi F, \qquad a^\dagger G=S_\psi b S_\varphi G, \quad  b^\dagger G=S_\psi a S_\varphi G, 
\label{48}\en
for all $F\in \Lc_\varphi$ and  $G\in \Lc_\psi$. Once again, following \cite{baginbagbook}, the operators $a$ and $b^\dagger$ could be called $S_\psi$-conjugate. A deeper investigation of these similarities conditions in the distributional sense could be interesting and it is in progress.

\section{Connection with the damped harmonic oscillator, preliminary results}\label{sectwpbs2}

In this section, using the analysis described in \cite{fff}, we show how a rigorous treatment of the quantized version of the system described by the Bateman lagrangian, \cite{bate},  suggests the use of wpbs.

The Bateman lagrangian is
\be L=m\dot x\dot y+\frac{\gamma}{2}(x\dot y-\dot xy)-kxy,
\label{51}\en
which returns the equations $m\ddot x+\gamma \dot x+kx=0$ and $m\ddot y-\gamma \dot y+ky=0$, where $m,\gamma$ and $k$ are the physical positive quantities of the oscillator. The first equation is associated to the damped harmonic oscillator (DHO), while the second to a virtual amplified oscillator, which gain what is lost by the first. The conjugate momenta are $$p_x=\frac{\partial L}{\partial \dot x}=m\dot y-\frac{\gamma}{2}\,y,\qquad
p_y=\frac{\partial L}{\partial \dot y}=m\dot x+\frac{\gamma}{2}\,y,
$$
and the corresponding classical Hamiltonian is
\be
H=p_x\dot x+p_y \dot y-L=\frac{1}{m} p_xp_y+\frac{\gamma}{2m}\left(yp_y-xp_x\right)+\left(k-\frac{\gamma^2}{4m}\right)xy.
\label{52}\en
By introducing the new variables $x_1$ and $x_2$ through 
\be
x=\frac{1}{\sqrt{2}}(x_1+x_2), \qquad y=\frac{1}{\sqrt{2}}(x_1-x_2),
\label{53}\en
$L$ and $H$ can be written as follows:
$$
L=\frac{m}{2}(\dot x_1^2-\dot x_2^2)+\frac{\gamma}{2}(x_2\dot x_1-x_1\dot x_2)-\frac{k}{2}(x_1^2-x_2^2)
$$
and
$$
H=\frac{1}{2m}\left(p_1-\frac{\gamma}{2}x_2\right)^2-\frac{1}{2m}\left(p_2-\frac{\gamma}{2}x_1\right)^2+\frac{k}{2}(x_1^2-x_2^2),
$$
where $p_1=\cfrac{\partial L}{\partial \dot x_1}=m\dot x_1+\cfrac{\gamma}{2}\,x_2$ and $p_2=\cfrac{\partial L}{\partial \dot x_2}=m\dot x_2-\cfrac{\gamma}{2}\,x_1$. We introduce next $\omega^2=\cfrac{k}{m}\,-\cfrac{\gamma^2}{4m^2}$, which we assume here to be strictly positive\footnote{The case $\omega^2<0$ is considered in \cite{fff}, where it is also stated that the case $\omega^2=0$ must be analysed with different techniques.}. We can rewrite $H$ as follows:

\be
H=\left(\frac{1}{2m}p_1^2+\frac{1}{2}m\omega^2x_1^2\right)-\left(\frac{1}{2m}p_2^2+\frac{1}{2}m\omega^2x_2^2\right)-\frac{\gamma}{2m}(p_1x_2+p_2x_1).
\label{54}\en

Following \cite{nakano} we impose the following canonical quantization rules between $x_j$ and $p_k$: $[x_j,p_k]=i\delta_{j,k}\1$, working in unit $\hbar=1$. Here $\1$ is the identity operator. This is equivalent to the choice in \cite{fesh}. Ladder operators can now be easily introduced:
\be a_k=\sqrt{\frac{m\omega}{2}}\,x_k+i\sqrt{\frac{1}{2m\omega}}\,p_k,
\label{55}\en
$k=1,2$. These are bosonic operators since they satisfy the canonical commutation rules: $[a_j,a^\dagger_k]=\delta_{j,k}\1$. 
In terms of these operators the quantum version of the Hamiltonian $H$ in (\ref{54}) can be written as
\be
	H=\omega\left(a_1^\dagger a_1-a_2^\dagger a_2\right)+\cfrac{i\gamma}{2m}\left(a_1a_2-a_1^\dagger a_2^\dagger\right)
\label{56}\en 

Following \cite{nakano}, it is possible to rewrite $H$ in a diagonal form by introducing
\be
A_1=\frac{1}{\sqrt{2}}(a_1-a_2^\dagger), \quad A_2=\frac{1}{\sqrt{2}}(-a_1^\dagger+a_2),
\label{57}\en
as well as 
\be
B_1=\frac{1}{\sqrt{2}}(a_1^\dagger+a_2), \quad B_2=\frac{1}{\sqrt{2}}(a_1+a_2^\dagger).
\label{58}\en
These operators satisfy the following requirements:
\be
[A_j,B_k]=\delta_{j,k}\1\\,
\label{59}
\en
with $B_j\neq A_j^\dagger$, $j=1,2$.
In fact, in terms of these operators, $H$ can now be written as follows:
\be
H=\omega\left(B_1A_1-B_2A_2\right)+\cfrac{i\gamma}{2m}\left(B_1A_1+B_2A_2+\1\right),
\label{510}\en 
which only depends on the pseudo-bosonic number operators $N_j=B_jA_j$.  The next proposition, proven in \cite{fff}  shows why  wpbs are useful to deal with the DHO.

\begin{prop}\label{prop1}
	There is no non-zero function $\varphi_{00}(x_1,x_2)$ satisfying $$A_1\varphi_{00}(x_1,x_2)=A_2\varphi_{00}(x_1,x_2)=0.$$ Also, there is no non-zero function $\psi_{00}(x_1,x_2)$ satisfying $$B_1^\dagger\psi_{00}(x_1,x_2)=B_2^\dagger\psi_{00}(x_1,x_2)=0.$$
\end{prop} 

The key of the proof, see \cite{fff}, is that the  solution of  $A_1\varphi_{00}(x_1,x_2)=A_2\varphi_{00}(x_1,x_2)=0$ must be of the form $\varphi_{00}(x_1,x_2)=\alpha\delta(x_1-x_2)$, $\alpha\in\mathbb{C}$. Analogously, $B_1^\dagger\psi_{00}(x_1,x_2)=B_2^\dagger\psi_{00}(x_1,x_2)=0$ only if  $\psi_{00}(x_1,x_2)=\beta\delta(x_1+x_2)$, $\beta\in\mathbb{C}$.

The situation is completely analogous to that described in Section \ref{sectafa}. The only difference is that we are here forced to consider a two-dimensional version of wpbs. This is almost automatic, and no extra problem is expected. A full analysis of this connection will be undertaken soon.

\section{Conclusions}\label{sectconcl}

Motivated by a recent result on the DHO we have introduced here a particular version of pseudo-bosons, which we have called {\em weak} since it is naturally defined in the space of distributions. We have shown what can be extended to this new situation, and we have analyzed in detail the weak pseudo-bosonic operators arising from the position  and from the  space derivative operators, which do not admit any square-integrable vacuum. We have shown that a similar framework as the one constructed for $\D$-pseudo bosons can be found also for wpbs. However, some of the good properties we had in the first case, are not guaranteed for these last operators. In particular, we don't know if the (generalized) eigenstates $\varphi_n$ and $\psi_n$ are, or are not, quasi bases on some dense domain in $\Hil$. However, at least in the concrete example considered in this paper, they turn out to be quasi-bases in a rather {\em large} space of functions.

The next step of our research will be to carry on the analysis of the DHO in terms of wpbs, and to see if other dissipative systems can be successfully described using these (or related) operators.

\section*{Acknowledgements}

The author acknowledges partial support from Palermo University and from G.N.F.M. of the INdAM.

\end{document}